\documentclass{osa-article}
\usepackage{braket}
\usepackage{lineno}
\usepackage[utf8]{inputenc}
\usepackage[T1]{fontenc}
\usepackage{babel}
\journal{osac}
\articletype{Research Article}
\begin{document}

\title{Type of Non-reciprocity in Fiber Sagnac Interferometer induced by Geometric Phases}

\author{Dongzi Zhao,\authormark{1} Jing-Zheng Huang,\authormark{1} Tailong Xiao,\authormark{1} Hongjing Li\authormark{1}, Xiaoyan Wu and Guihua Zeng\authormark{1}}

%\address{\authormark{1}School of Electronic Information and Electrical Engineering, State Key Laboratory of Advanced Optical Communication Systems and Networks, Center of Quantum Sensing and Information Processing, Shanghai Jiao Tong University, Shanghai 200240, China}
\address{\authormark{1}State Key Laboratory of Advanced Optical Communication Systems and Networks, Institute of Quantum Sensing and Information Processing, Shanghai Jiao Tong University, Shanghai 200240, China}

%\email{\authormark{*}jzhuang1983@sjtu.edu.cn}

%%%%%%%%%%%%%%%%%%% abstract %%%%%%%%%%%%%%%%

\begin{abstract}
    %The non-reciprocity offers the Sagnac interferometer a wide range of applications, but at the same time limits its accuracy.  
    %Therefore, research on it has always been in an important state. 
    %Therefore, 
    %for the research work of SI, 
    %the study of non-reciprocity has always been the focus of Sagnac interferometer research.
    %On the one hand, the Sagnac interferometer is widely used in the sensor field, because any physical quantity introducing non-reciprocity could be sensed by it. On the other hand, due to the variety of non-reciprocal factors, disturbance may occur in various way. 
    %By taking PDL into consideration we found a new type of non-reciprocal phase existed in Sagnac interferometer, which is not caused by optical path difference but by the evaluation of polarization state.
    The non-reciprocity of Sagnac interferometer provides ultra-high sensitivity for parameter estimation and offers a wide range of applications, especially for optical fiber sensing.
    In this work, we study a new type of non-reciprocity existed in optical fiber Sagnac interferometer where the polarization dependent loss is taken into consideration. 
    In particular, this non-reciprocity is irrelevant to the physical effects that being considered in previous studies, which originates from the geometric phases induced by continuous-weak-measurement. In consequence, it has a unique phenomenon of sudden phase transition, which may open a new way for the future design of high precision optical fiber sensors.
    %This newly found non-reciprocity is different from previous ones in that it is caused by the evolution of the polarization state of light and its unique phase transition characteristic can be explained by the theory of geometric phase. 
    %This non-reciprocal phase is found to have properties similar to the geometric phase in quantum mechanics theory.【1.强调其不同于以往研究的非互易性；2.该非互易性可以用几何相位解释；3.shed a new light for future high precision sensor design】
\end{abstract}

%%%%%%%%%%%%%%%%%%%%%%%%%%  body  %%%%%%%%%%%%%%%%%%%%%%%%%%
\section{Introduction}
    %The fiber Sagnac interferometer plays important roles in various applications, the most important and famous one is the fiber optic gyroscope (FOG) based on Sagnac effect\cite{ref1,ref2}. Moreover, it detects not only rotation rate but any non-reciprocal phenomena that occur as the light traverses the optic fiber coil. Consequently, time varying physical effects can also be detected, which imposes many other applications such as current detection\cite{ref3}, temperature sensing\cite{ref4}, strain detectors\cite{ref5} and so on\cite{ref6}.
    
    The optical fiber Sagnac interferometer plays important roles in various applications, such as fiber optic gyroscope (FOG)\cite{ref1,ref2}, current detection\cite{ref3}, temperature sensing\cite{ref4}, strain detectors\cite{ref5} and so on\cite{ref6}. Standing in the heart of all these applications, it is the non-reciprocity of Sagnac loop induced by varying physical effects. However, the versatility of Sagnac inteferomter sometimes becomes a curse for one to pursue high accuracy and precision. For example, in the case of FOG, the irrelevant effects caused by magnetic fields, thermal transits, acoustic fields and non-linearity of optic fiber\cite{ref7,ref8} will all induce undesired non-reciprocal phases, which dramatically limit the performance of the FOG and make constrains on the design of optic fiber coil to minimize these irrelevant effects\cite{ref6}. 
    
    In this work, a new type of non-reciprocity exists in fiber Sagnac interferometer is studied. Different from the previous studies, the non-reciprocal phase is irrelevant to the physical effects we mentioned above, but because the optical axes of the birefringence and polarization dependent loss (PDL) in the optic fiber are in different directions. Moreover, when the PDL strength reaches a certain level, there will be a sudden transition of this non-reciprocal phase. In the beginning, the phenomenon is proposed and analyzed based on the traditional transfer matrix method (TMM)\cite{ref20,ref19}. To further explore its origin, we employ a recent proposed continuous-weak-measurement model\cite{ref18}, and reveal its deep relation to the geometric phase. 

    %
    %In general, the non-reciprocal phases are caused by the optical path difference introduced by various factors. Consider birefringence and polarization dependent loss (PDL) in the optic fiber with different axes, there would be a phase difference between the light propagating in the forward and reverse directions of the Sagnac intereformeter, which means there is non-reciprocity. This phase difference has the following characteristics: it has nothing to do with the path but is related to the angle between the birefringence axes and the PDL axes and the strength of PDL. When PDL reaches a certain value, phase transition would happen.Previous work has mentioned a method to analyze fiber by dividing it into small sections with birefringence axes twisted between each other\cite{ref19}.By expressing each small section with Jones matrix, the influence of the entire fiber could be obtained.This analytical method can get numerical conclusions, but it cannot physically explain the cause of this phenomenon.We have conducted further research on this and found that this phenomenon can be explained by the geometric phase theory in quantum mechanics. 
    
    This paper is organized as follows. The phenomenon of a new type non-reciprocity in Sagnac interferometer is presented along with a theoretical analysis based on TMM in Sec.II. Furthermore, the physical insight of this non-reciprocity is explored by using the model of continuous weak measurement and geometric phase in Sec.III, and the corresponding influences (both negative effects and potential for sensing) in the context of optical sensing are discussed Sec.IV. Finally, conclusions are made in Sec.V.

\section{Principle}
\begin{figure}[h!]
        \centering\includegraphics[width=9cm]{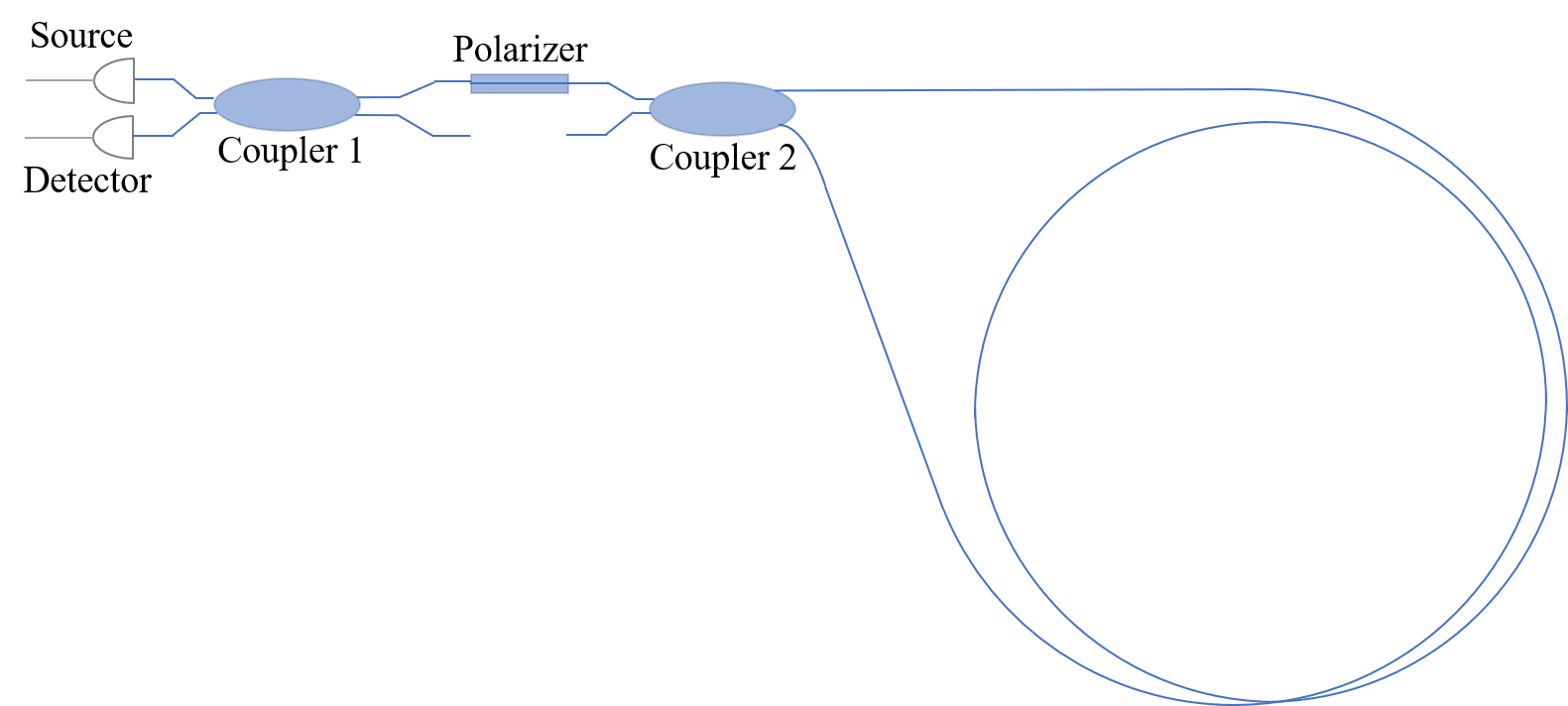}
        \caption{The Schematic diagram of Sagnac interferometer.}\label{Fig:sagnac_principle}
    \end{figure}

    %\subsection{Preliminary}
    The basic idea of a fiber Sagnac interferometer is shown in Fig.\ref{Fig:sagnac_principle}\cite{ref6}. The light ejected from a source passes through a beam splitter (Coupler 1) and a polarizer, and then divided into two beams by another beam splitter (Coupler 2), and then traverse along the clockwise (CW) and counterclockwise (CCW) directions of the fiber ring respectively. Consequently, any time varying physical effects and inherently non-reciprocal phenomena can produce a non-reciprocal phase-shift, which can be finally detected by the detector placed on the other end of Coupler 1.

    %\subsection{Non-reciprocity in fiber-loop induced by polarization dependent loss}
    %Previous analysis mainly focused the rotation of the optical fiber ring, or the influence of some time-varying parameters on the optical fiber ring, and some also took the birefringence into consideration.
    As a common feature in optical fiber, the PDL can be easily generated by a number of physical effects. For examples, the pressure in the diametrical direction of the fiber will make the transmission coefficients of the two orthogonal polarization states being different, along with extra birefringence causing by the elasto-optic effect\cite{ref9}.
    %will cause stress and strain in the fiber core. 
    %which will introduce birefringence due to the elasto-optical effect\cite{ref9}. 
    %When the pressure reaches a sufficient level, it will induce PDL, bacause of the change in transmission coefficient introduced by core deformation. 
    %due to the difference between the transmission coefficient of the stressed shaft and the unstressed shaft, it will further introduce PDL\cite{ref9}.
    Similarly, the bending of the optical fiber will cause extra birefringence and extra PDL at the same time\cite{ref10,ref11,ref12,ref13,ref14,ref15}. Although the PDL and the bireferingence simultaneously appear in the fiber, their principal axes may in general be not coincident \cite{Wang2007,not_align,bend1,bend2}.
    \begin{figure}[h!]
      \centering\includegraphics[width=6cm]{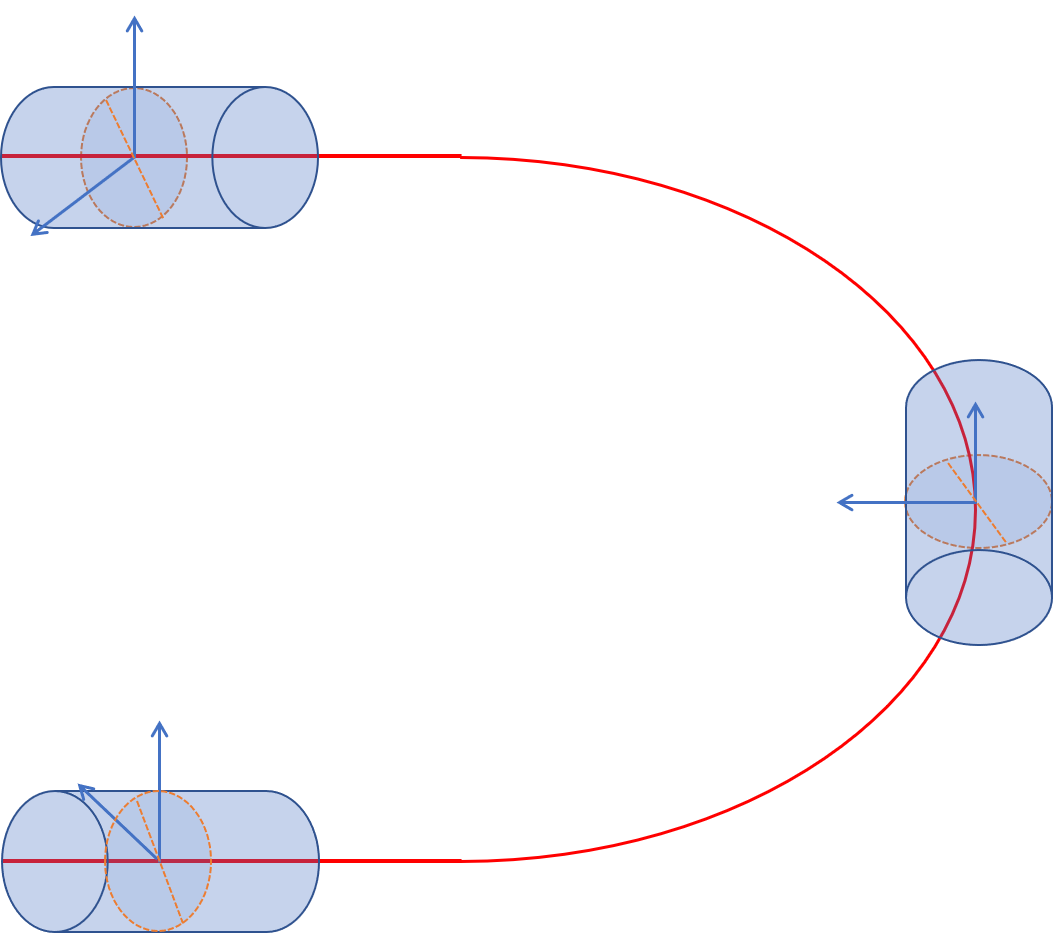}
        \caption{A simple situation where the principal axes of PDL and birefringence in the fiber are differed by a fixed angle with respect to each other.}\label{Fig:principal_axes_1}
    \end{figure}
    \begin{figure}[h!]
     \centering\includegraphics[width=6cm]{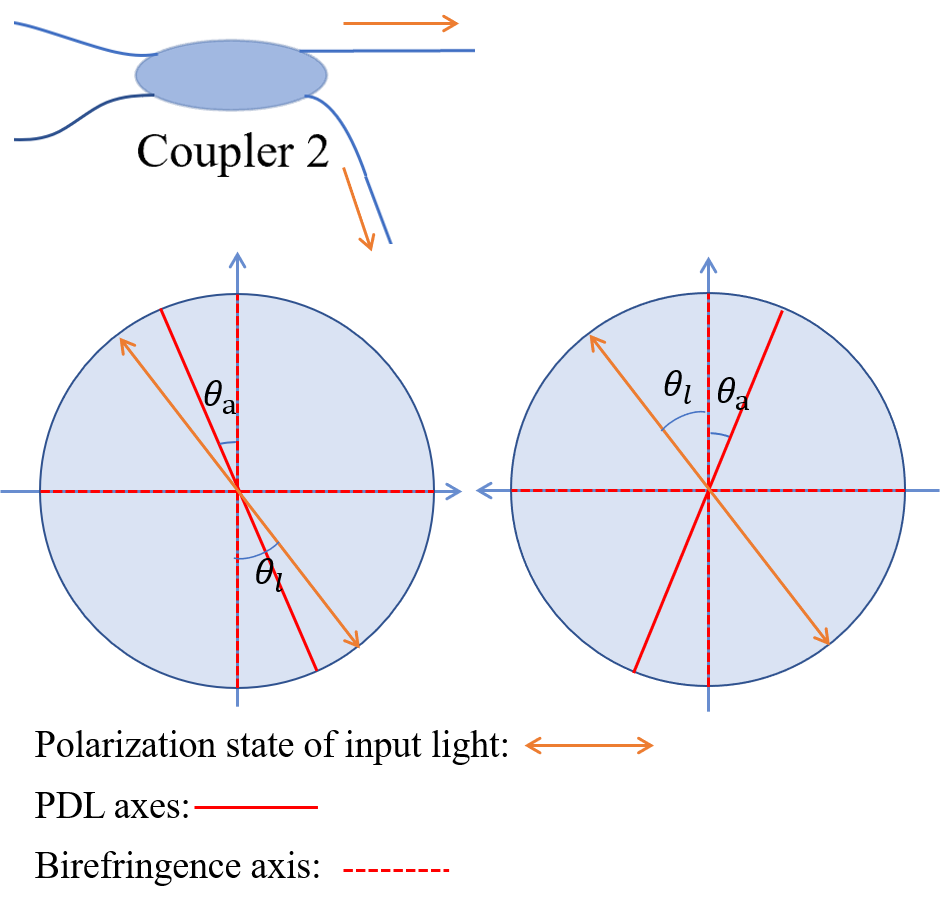}
        \caption{Due to the geometric characteristics of the optical fiber loop, the principal axes of PDL and birefringence in the fiber exhibits ‘symmetric’ characteristics at the two ports. Here, $\theta_a$ represents the angle between the PDL axis and the birefringence axis, and $\theta_l$ represents the angle between liner input polarization and the birefringence axis.}\label{Fig:principal_axes_2}
    \end{figure}
    For simplicity, we assume the principal axes of the birefringence and PDL of the fiber loop in Fig.\ref{Fig:sagnac_principle} are fixed and differed by an angle $\theta_a$ with respect to each other, as is shown in Fig.\ref{Fig:principal_axes_1} and Fig.\ref{Fig:principal_axes_2}.
    %Then, since the optical axes of the birefringence and PDL always have a fixed angle with the local coordinate system, they will show a symmetrical relationship at the two ports, as shown in Fig.2. 
    %
    %When the optical axes of birefringence does not coincide with the local coordinate system, the situation will be transformed into the problem of the birefringence introduced optical path difference[].
    %
    Here, we make the principal axes of birefrigence as the laboratory coordinate. 
    %In general,the axes of birefrigence and PDL are in different directions.
    %Based on the above assumptions, we can carry out further analysis.
    To analyze the evolution of optical polarization when the light traverses through the fiber loop, we divide the fiber loop into a series of small segments and employ the transfer matrix method (TMM)\cite{ref19}. First, the optical fiber loop is divided into N segments, and the length ($l$) of each segment is made much smaller than the beat length of the optical fiber. We can then consider each of the segments as a combination of a birefringent wave plate and a non-ideal polarizer, as is depicted in Fig.\ref{Fig:fiber_segment}. 
    The effective birefringent wave plate and non-ideal polarized of the $k$th segment can be described by Jones matrices
        $\left[
            \begin{array}{cc}
                1&0\\
                0&e^{-i\phi_k}\\
            \end{array}
        \right]$
         and 
        $\left[
            \begin{array}{cc}
                1&0\\
                0&\sqrt{\eta_k}\\
            \end{array}
        \right]$
    respectively, with the overall phase shift and transmittance being omitted for simplicity.
    Here, $\phi_k=2\pi(n_o-n_e)l/\lambda$ is the phase difference introduced by the effective birefringent wave plate, with $\lambda$ being the wavelength of the light, $n_o$ and $n_e$ being the refractive indexes along the two orthogonal principal axes.
    $\eta = T_{min}/T_{max} \approx 1$ is the PDL strength of the effective non-ideal polarizer, of which the polarization-dependent transmittance being denoted by $T_{max}$ and $T_{min}$ respectively.
    %is the ratio of the main transmittance and the minimum transmittance of the non-ideal polarizer, 
    %$\sqrt{\eta}$ is the PDL strength of the effective non-ideal polarizer,
    %It also characterizes the PDL strength. The closer $\eta$ is to 1, the smaller the PDL strength, and vice versa. 
    In consequence, the overall PDL strength of the optical fiber loop is given by
    \begin{equation}
        \Gamma_{dB} = -10\log\eta^N = -10N\log\eta.
    \end{equation}
    \begin{figure}[h!]
        \centering\includegraphics[width=9cm]{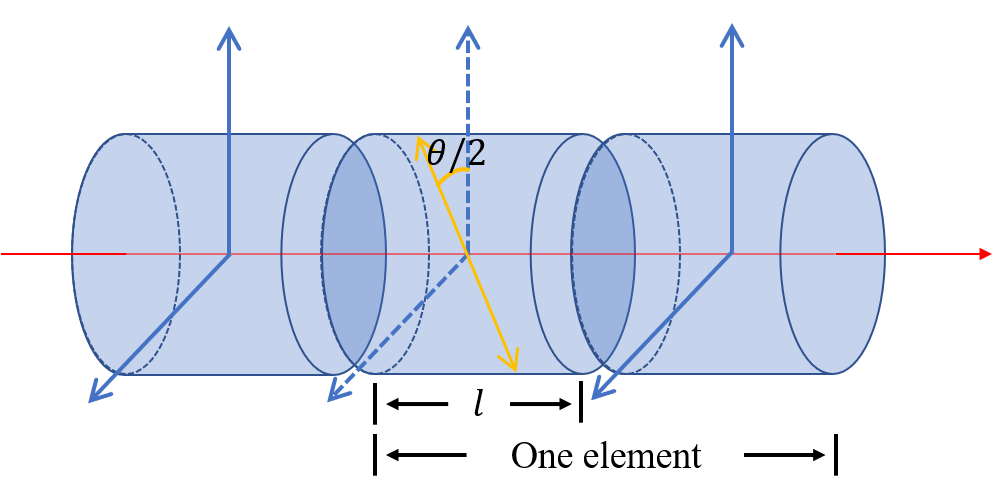}
        \caption{A divided segment of the optical fiber.}\label{Fig:fiber_segment}
    \end{figure}
    The Jones matrix of the $k$th segment can be expressed in the following form:
        \begin{equation}
        \begin{array}{c}
        \left[
            \begin{array}{cc}
                cos(\theta_k/2)&\!sin(\theta_k/2)\\
                sin(-\theta_k/2)&\!cos(\theta_k/2)\\
            \end{array}
        \right]
        \left[
            \begin{array}{cc}
                1&\!\!0\\
                0&\!\!\sqrt{\eta_k}\\
            \end{array}
        \right]
        \left[
            \begin{array}{cc}
                cos(\theta_k/2)&\!sin(-\theta_k/2)\\
                sin(\theta_k/2)&\!cos(\theta_k/2)\\
            \end{array}
        \right]
        \left[
            \begin{array}{cc}
                1&\!\!0\\
                0&\!\!e^{-i\phi_k}\\
            \end{array}
        \right]
        \\=R^{-1}(\theta_k)M_{\eta_k}R(\theta_k)C(\phi_k),    
        \end{array},
        \end{equation}
    where $\theta_k/2$ is the angle between the principal axes of the effective non-ideal polarizer and effective phase plate.
        
    We then consider an input light with liner polarization state of $\ket{\psi_{in}}=\left[\begin{array}{c}cos(\theta/2)\\sin(\theta/2)\\ \end{array}\right]$. 
    %The optical fiber loop is divided into N segments, where the length of each segment is much smaller than the beat length of the optical fiber. 
    %For the $k$th segment, its Jones matrix is $R^{-1}(\theta_k)M_{\eta_k}R(\theta_k)C(\phi_k)$.
    %we define the Jones matrix of the birefringent crystal as: $C(\psi_k)=\left[\begin{array}{cc} 1&0\\ 0&e^{-i\psi_k}\\ \end{array}\right]$, and define the Jones matrix of the non-ideal polarizer as
    %$R^{-1}(\theta_k)M_{\eta_k}R(\theta_k)$, where $R(\theta_k)=\left[\begin{array}{cc} cos(\theta_k)&sin(-\theta_k)\\
    %sin(\theta_k)&cos(\theta_k)\\
    %\end{array}\right]$,$M_{\eta_k}=\left[\begin{array}{cc}1&0\\0&\eta_k\\ \end{array}\right]$.
    For simplicity, we assume that the axes and strengths of birefringence and PDL are time-invariant, which means $\theta_{k}=\theta$, $\eta_{k}=\eta$, $\phi_{k}=\phi$. Here we assume that PDL axis and input polarization are aligned, so the same $\theta/2$ is used to indicate the direction of the PDL axis and the input state. For the light propagating in the CW direction, the transfer matrix of the fiber loop evolves to be:
    \begin{equation}
        \ket{\psi_{out+}}=\prod{R^{-1}(\theta)M_{\eta}R(\theta)C(\phi)\ket{\psi_{in}}}.
    \end{equation}
    %$$\ket{\psi_{out+}}=\prod{R^{-1}(\theta)M_{\eta}R(\theta)C(\psi)\ket{\psi_{in}}},\eqno{(2)}$$
    On the other hand, the light propagating in the CCW direction evolves to be:
    \begin{equation}
        \ket{\psi_{out-}}=\prod{C(\phi)R(\theta)M_{\eta}R^{-1}(\theta)\ket{\psi_{in}}}.
    \end{equation}
    %$$\ket{\psi_{out-}}=\prod{C(\psi)R(\theta)M_{\eta}R^{-1}(\theta)\ket{\psi_{in}}},\eqno{(3)}$$
    By projecting both $\ket{\psi_{out+}}$ and $\ket{\psi_{out-}}$ back to the initial polarization state $\ket{\psi_{in}}$, the phase difference between the CW light and the CCW light can be obtained by:
    \begin{equation}
        \braket{\psi_{out+}|\psi_{in}}=\sqrt{I_+}e^{i\phi_+},\braket{\psi_{out-}|\psi_{in}}=\sqrt{I_-}e^{i\phi_-}.
    \end{equation}
    %
    %$$\braket{\psi_{out+}|\psi_{in}}=\sqrt{I_+}e^{i\psi_+},\braket{\psi_{out-}|\psi_{in}}=\sqrt{I_-}e^{i\psi_-},\eqno{(4)}$$
    
\begin{figure}[h!]
        \centering\includegraphics[width=10cm]{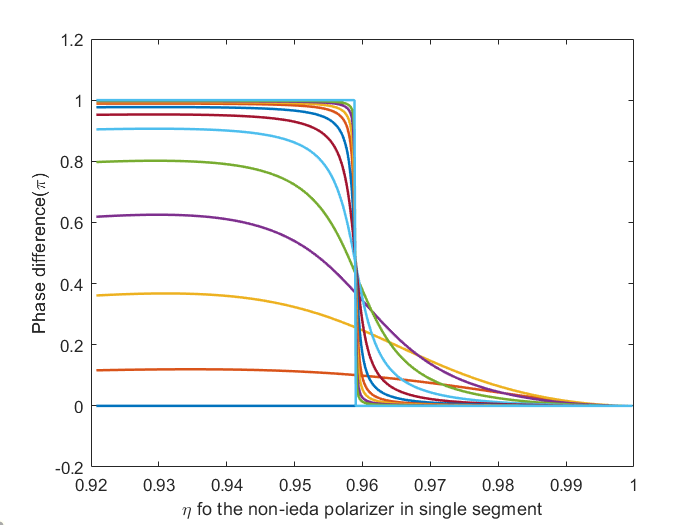}
        \caption{Sorted by the left end of the curve, from bottom to top in the figure are the phase difference changes with $\eta$ when $\theta=0, \pi/4, 3\pi/8, 7\pi/16, 15\pi/32, 31\pi/64, 63\pi/128,$ $127\pi/256, 255\pi/512$, $511\pi/1024, 1023\pi/2048, 2047\pi/4096, \pi/2$. It can be seen that as $\theta$ approaches $\pi/2$, the slope of the curve neer $\eta = \eta_c$ will gradually increase until a sudden phase transition occurs, where $\eta_c\approx 0.96$. The variation range of $\eta$ in the figure corresponds to the variation range of $\Gamma_{dB}$ from 0dB to 84.56dB. At the transition point, $\Gamma_{dB}\approx 42dB$. }\label{Fig:phase_diff_2d}
    \end{figure}

    The change of phase difference $\phi_+-\phi_-$ along with $\eta$ by choosing specific values of $\theta$ is numerically simulated in Fig.\ref{Fig:phase_diff_2d}. Here we choose $N=200$, $\psi=2\pi/N$, the variation range of $\eta$ is from 0.92 to 1, and the corresponding variation range of $\Gamma_{dB}$ is from 84.56dB to 0dB.  According to the numerical simulation results, we find that a phase transition appears when $\eta$ reaches a certain value. 
    \section{Theory}
    \subsection{Theoretical analysis based on measurement-induced geometric phase}
    With assistance of TTM, we have discovered a new type of non-reciprocal phase induced by PDL, but its physical essence is not clear. In order to explain this phenomenon, we introduce a measurement-induced geometric phase theory that recently proposed by Gebhart, et al.\cite{ref18}. In this theory, it was shown that Geometric phases are not necessarily a consequence of adiabatic time evolution, but can also be induced by a cyclic sequence of quantum measurements. Moreover, the mapping between the measurement sequence and the geometric phase undergoes a topological transition by varying the measurement strength\cite{ref18}. The $k$th variable strength measurement can be written by the follow form:
    %In this theory, the system state $\ket{\psi_0}$ under going a sequence of weak measurement and in the end projected to the initial state would accumulate a geometric phase. In this theory, each measurement process  can be expressed as follows:
    \begin{equation}
        M_k^{r_k}(\boldsymbol n_k,\eta_k)
        =\boldsymbol R^{-1}(\boldsymbol n_k)M_k^{r_k}(\boldsymbol e_z,\eta_k)\boldsymbol R(\boldsymbol n_k),
    \end{equation}
    where $r_k\in\{+,-\}$ is the measurement readout, $\eta_k$ is the measurement strength, and
    \begin{equation}
        M_k^{+}(\boldsymbol e_z,\eta_k)=\left[
          \begin{array}{cc}
           1 & 0\\
           0 & \sqrt{\eta_k}\\
          \end{array}
    \right],
    M_k^{-}(\boldsymbol e_z,\eta_k)=\left[
          \begin{array}{cc}
           0 & 0\\
           0 & \sqrt{1-\eta_k}\\
          \end{array}
    \right]
    \end{equation}
    % $$
    % M_k^{+}(\boldsymbol e_z,\eta_k)=\left[
    %       \begin{array}{cc}
    %       1 & 0\\
    %       0 & \sqrt{\eta_k}\\
    %       \end{array}
    % \right],
    % M_k^{-}(\boldsymbol e_z,\eta_k)=\left[
    %       \begin{array}{cc}
    %       0 & 0\\
    %       0 & \sqrt{1-\eta_k}\\
    %       \end{array}
    % \right]
    % $$
    are the operators of measurement on z direction, corresponding to the readouts of $+$ and $-$ respectively. And the operator
    %$$\boldsymbol n_k=(sin(\theta_k)cos(\psi_k),sin(\theta_k)sin(\psi_k),cos(\theta_k))\eqno{(6)}$$
    \begin{equation}
        R(\boldsymbol n_k)=\left[
          \begin{array}{cc}
           cos(\theta_k/2)&e^{-i\phi_k}sin(\theta_k/2)\\
           sin(-\theta_k/2)&-e^{-i\phi_k}cos(\theta_k/2)\\
          \end{array},
    \right]
    \end{equation}
    % $$R(\boldsymbol n_k)=\left[
    %       \begin{array}{cc}
    %       cos(\theta_k/2)&e^{-i\phi_k}sin(\theta_k/2)\\
    %       sin(-\theta_k/2)&-e^{-i\phi_k}cos(\theta_k/2)\\
    %       \end{array}
    % \right]$$ 
    rotates the measurement orientation from $z$ to $n_k=(sin(\theta_k)cos(\phi_k),sin(\theta_k)sin(\phi_k),cos(\theta_k))$.
    %, $r_k=+,-$ is the two possible readouts. In this paper,we only consider the state when $r_k = +$, because light in fiber could always pass through the segment with low PDL strength. Then
    Interestingly, the operator $M_k^{+}(\boldsymbol n_k,\eta_k)$ can be effectively implemented by rotating an imperfect polarizers and adding phase plate\cite{ref18}, which is perfectly matches the optical fiber segment model we have proposed in the last section. In particular, the PDL matrix $M_{\eta}$ is exactly the same with $M_k^{+}(\boldsymbol e_z,\eta_k)$ if we consider the PDL strength as the measurement strength. 
    
\begin{figure}[h!]
        \centering\includegraphics[width=10cm]{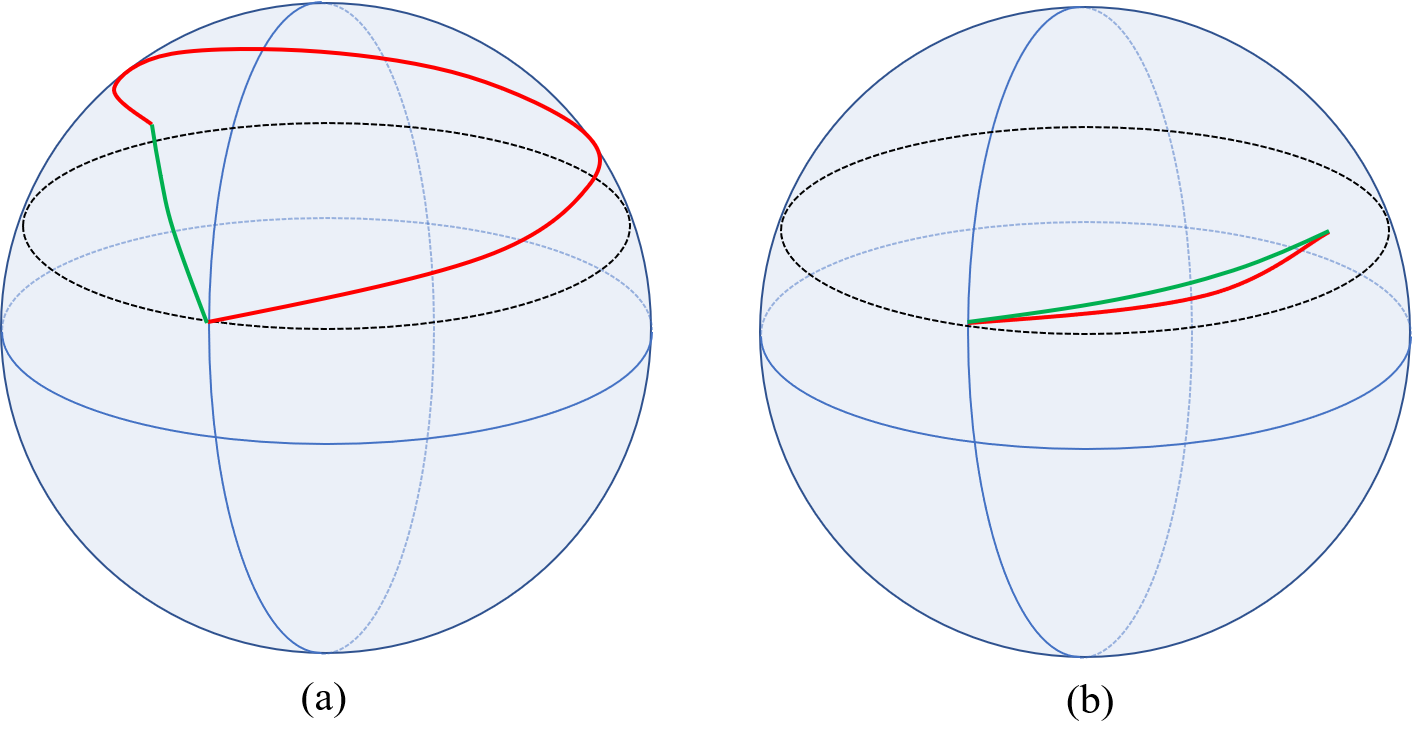}
        \caption{When taking $\theta_k=\pi/4$ and $\phi_k=2\pi/N$, the state trajectories under different measurement strength are depicted on the Bloch sphere. The circle drawn by the black dashed line represents the measurement orientation of the measurement sequences (\{$n_k$\}). The red curves in (a) and (b) represent the state trajectories under different measurement strength. And the green curves represent the final projection process. We can find that the solid angle of trajectory may be vary different under different measurement strength.
        }\label{Fig:Bloch_2}
    \end{figure}

    Therefore, the process described in Sec.II can be described a quasicontinuous measurement sequence with postselecting the results of $r_k=+$. According to the analysis made in Ref.\cite{ref18}, after $N$ quasicontinuous measurements the initial state $\ket{\psi_in}$ becomes:
    \begin{equation}
        \ket{\psi_N}=M^{+}_N(\boldsymbol n_N,\eta_N) M^{+}_{N-1}(\boldsymbol n_{N-1},\eta_{N-1})...M^{+}_1(\boldsymbol n_1,\eta_1)\ket{\psi_{in}}.
    \end{equation}
    Different measurement sequences induce state trajectories on the Bloch sphere. For example, by taking $\theta_k=\pi/4$ and $\phi_k=2\pi k/N$, the state trajectories under different measurement strength are depicted, as is shown in Fig.\ref{Fig:Bloch_2}. 
    %the trajectory forms a circle, as is shown in Fig.6. 
    If $\ket{\psi_N}$ is then projected back to $\ket{\psi_{in}}$, the corresponding geometric phase induced by this process is $\chi_{geom} = arg(\braket{\psi_{in}|\psi_N})$.
    According to the geometric phase theory, this measurement-induced geometric phase can be equivalently expressed via the solid angle $\Omega$ of the trajectory on Bloch sphere by $\chi_{geom} = \Omega/2$\cite{ref16,ref17}.
    
    Back to our analysis of the fiber Sagnac loop. Let us split the effective phase plate in the fiber segment by two parts (as is shown in Fig.\ref{Fig:Bloch_2}), and then $C(\phi_k)$ in Eq.(2) can be rewritten as
    \begin{equation}
    C(\phi_k)=
    \left[
    \begin{array}{cc}
        1 & 0 \\
        0 & e^{-i\phi_{k+1}}
    \end{array}
    \right]
    \left[
    \begin{array}{cc}
        1 & 0 \\
        0 & e^{i\phi_{k}}
    \end{array}
    \right]=C(\phi_{k+1})C^{-1}(\phi_k),
    \end{equation}
    % $$C(\phi_k)=
    % \left[
    % \begin{array}{cc}
    %     1 & 0 \\
    %     0 & e^{-i\phi_{k+1}}
    % \end{array}
    % \right]
    % \left[
    % \begin{array}{cc}
    %     1 & 0 \\
    %     0 & e^{i\phi_{k}}
    % \end{array}
    % \right]=C(\phi_{k+1})C^{-1}(\phi_k),$$
    where $\phi_{k+1}-\phi_k=\phi$.
    %\end{equation}
    %
    %$$
    %\left[
    %\begin{array}{cc}
    %    1 & 0 \\
    %    0 & e^{-i2\pi/N}
    %\end{array}
    %\right]
    %=
    %\left[
    %\begin{array}{cc}
    %    1 & 0 \\
    %    0 & e^{-i\psi_k}
    %\end{array}
    %\right]
    %\left[
    %\begin{array}{cc}
    %    1 & 0 \\
    %    0 & e^{i\psi_{k-1}}
    %\end{array}
    %\right]\eqno{(11)}
    %$$
    Then Eq.(3) and Eq.(4) can be rewritten as:
    \begin{equation}
    \begin{array}{lll}
        \ket{\psi_{out}}
        &=...C(\phi_{k+2})C^{-1}(\phi_{k+1})R^{-1}(\theta_{k+1})M_{\eta_{k+1}}R(\theta_{k+1})C(\phi_{k+1})\\&C^{-1}(\phi_{k})R^{-1}(\theta_{k})M_{\eta_k}R(\theta_{k})C(\phi_{k})C^{-1}(\phi_{k-1})...\ket{\psi_{in}}\\
        &=\sum C^{-1}(\phi_k)R^{-1}(\theta_k)M_{\eta_k}R(\theta)C(\phi_{k})\ket{\psi_{in}}.
    \end{array}
    \end{equation}
    %$$...C(\psi)R^{-1}(\theta)M_{\eta}R(\theta)C(\psi)R^{-1}(\theta)M_{\eta}R(\theta)C(\psi)...\ket{\psi_{in}}\eqno{(12)}$$
    %Then, according to (11), every $C(\psi)$ could be decomposed into $C(\psi_k)C^{-1}(\psi_{k-1})$. The whole process can be expressed as follow:
    %\begin{equation}
    %\begin{aligned}
    %    ...C(\psi_k)C^{-1}(\psi_{k-1})R^{-1}(\theta)M_{\eta}R(\theta)C(\psi_{k-2})C^{-1}\\(\psi_{k-3})R^{-1}(\theta)M_{\eta}R(\theta)C(\psi_{k-4})C^{-1}(\psi_{k-5})...\ket{\psi_{in}}
    %\end{aligned}
    %\end{equation}
    %$$...C(\psi_k)C^{-1}(\psi_{k-1})R^{-1}(\theta)M_{\eta}R(\theta)C(\psi_{k-2})C^{-1}(\psi_{k-3})R^{-1}(\theta)M_{\eta}R(\theta)C(\psi_{k-4})C^{-1}(\psi_{k-5})...\ket{\psi_{in}}\eqno{(13)}$$
    %The new expression of the recombined single segment is $$C^{-1}(\psi_{k})R^{-1}(\theta_k)M_{\eta_k}R(\theta(\psi_{k}).$$ 
    Consequently, the transfer matrix connecting $\ket{\phi_{k-1}}$ and $\ket{\phi_k}$ is:
    \begin{equation}
    \begin{array}{c}
    C^{-1}(\phi_{k})R^{-1}(\theta_k)M_{\eta_k}R(\theta_k)C(\phi_{k})
    \\
    \hspace{-2mm}=\!\!\left[\!
      \begin{array}{cc}
        cos(\theta_k/2)&\!\!sin(\theta_k/2)\\
        e^{-i\phi_k}sin(\theta_k/2)&\!\!-e^{-i\phi_k}cos(\theta_k/2)\\
      \end{array}\!
    \right]\!
    \left[\!
      \begin{array}{cc}
        1&\!\!\!0\\
      0&\!\!\!\sqrt{\eta_k}\\
      \end{array}\!
    \right]\!
    \left[\!
      \begin{array}{cc}
      cos(\theta_k/2)&\!\!e^{-i\phi_k}sin(\theta_k/2)\\
      sin(-\theta_k/2)&\!\!-e^{-i\phi_k}cos(\theta_k/2)\\
      \end{array}\!
    \right]\!,
    \end{array}
    \end{equation}
    which is coincident with Eq.(6). Consequently, the non-reciprocal phase difference between the light traverse along opposite directions in the fiber Sagnac loop can be interpreted as a PDL-induced geometric phase.
    
    \begin{figure}[h!]
        \centering\includegraphics[width=7cm]{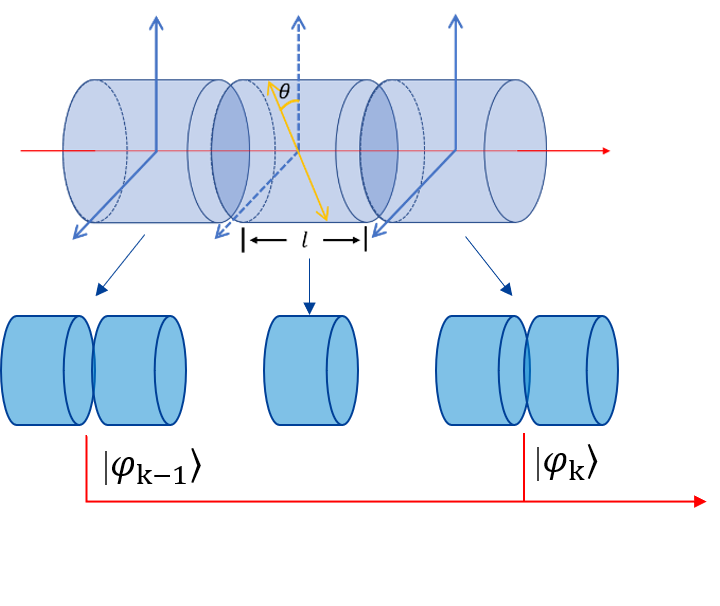}
        \caption{Splitting the birefringent wave plate in each segment into two parts.}
    \end{figure}\label{splitted_segment}
    
    \subsection{Numerical simulation}
    
    In order to study the PDL-induced non-reciprocity in the sense of geometric phase, we first recalculate the scenario that presented in Fig.\ref{Fig:phase_diff_2d}. By setting $\phi_k=2k\pi/N (k=1,2,3...N)$, $\theta_k = \theta$ (with $0<\theta<\pi$),  $\eta_k=\eta$ (with $1-\eta<<1$), 
    %This makes the change of the polarization state have a similar expression to the evolution of the quantum state above. 
    and making the initial polarization state as    $\ket{\psi_{in}}=\left[
    \begin{array}{c}
    cos(\theta/2)\\
    sin(\theta/2)
    \end{array}
    \right]$, 
    we can plot the polarization state trajectory on the Bloch sphere that similar to the one shown in Fig.\ref{Fig:Bloch_2}. 
    
    \begin{figure}[h!]
            \centering\includegraphics[width=10cm]{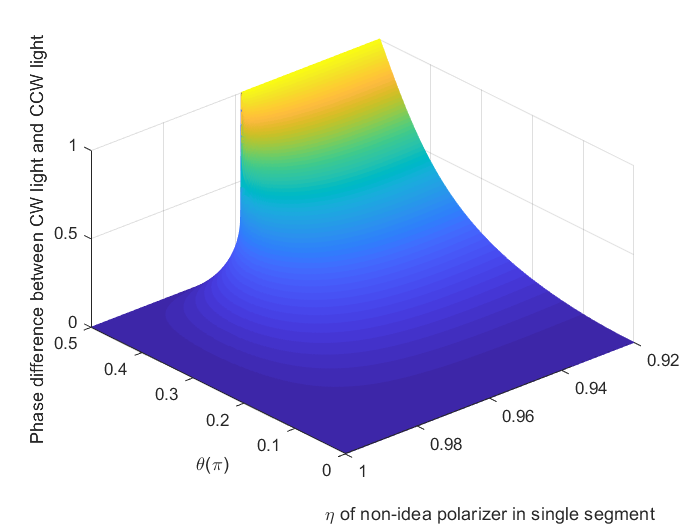}
            \caption{The phase difference between CW light and CCW light varies with $\theta$ and $\eta$.}\label{Fig:phase_diff}
    \end{figure}
    
    \begin{figure}[h!]
            \centering\includegraphics[width=8cm]{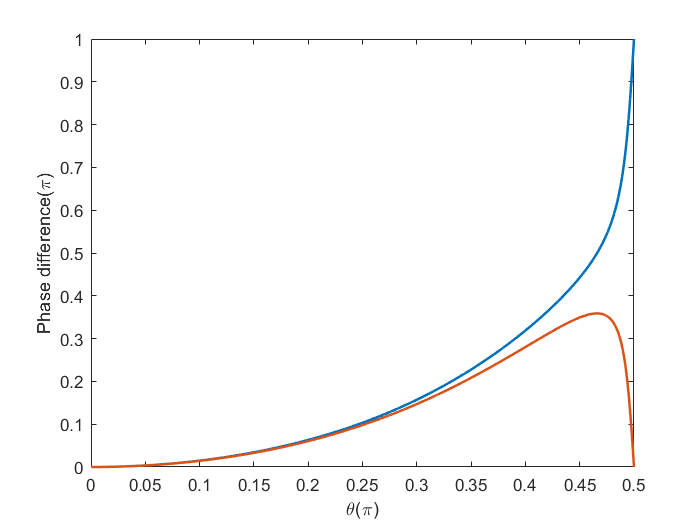}
            \caption{The curve of the phase change with $\theta$ before(red) and after(blue) the phase transition point.}\label{Fig:Phase_Tran}
    \end{figure}
    
    \begin{figure}[h!]
    \centering
      \includegraphics[width=13cm]{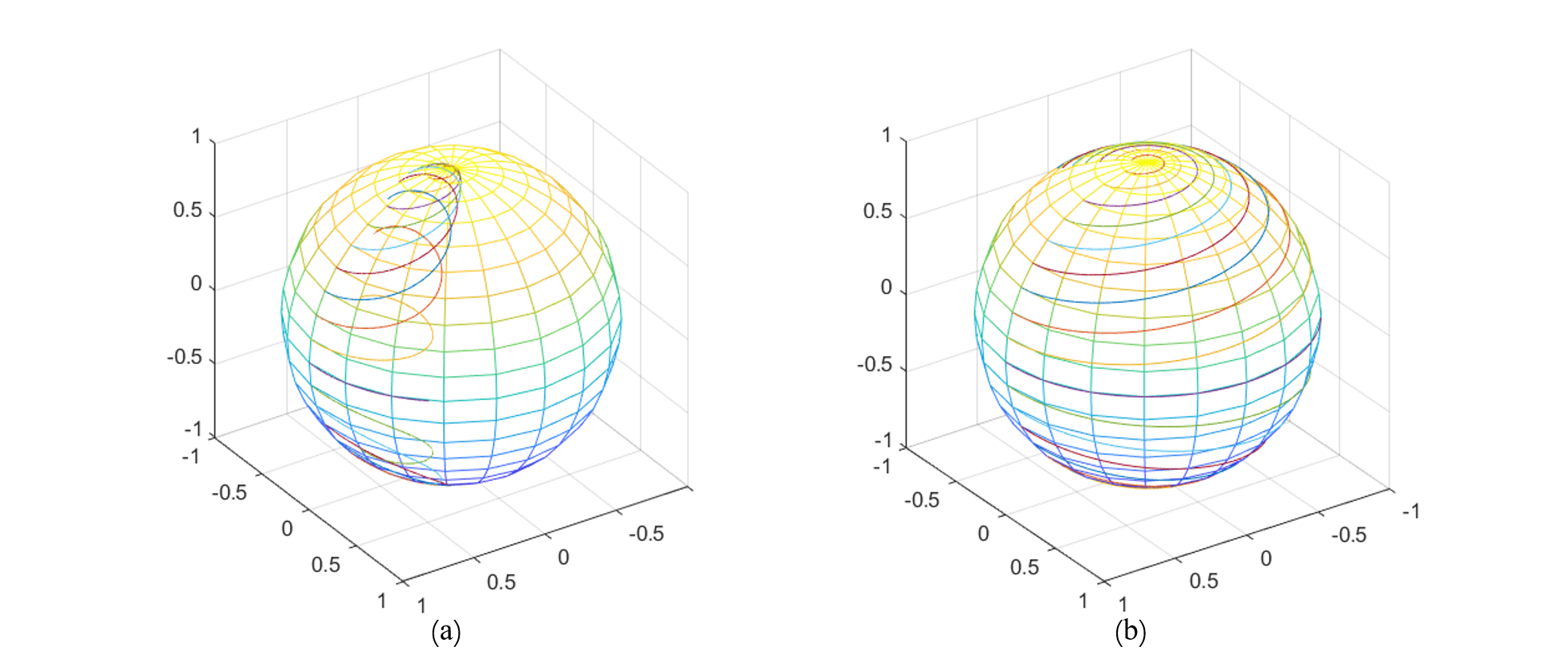} 
      \caption{The trajectory under different $\theta$ when $\eta>\eta_c$(a) and $\eta<\eta_c$(b).}\label{Fig:Bloch_3} 
    \end{figure}
    
    %Applying the theoretical techniques and results that proposed in Ref.\cite{ref18}, 
    Moreover, we find that the trajectory of polarization state on the Bloch sphere changes along with the PDL strength. Because the final projection will make the polarization state return to the initial state with the shortest trajectory on the Bloch sphere, as is shown in Fig.\ref{Fig:Bloch_2}.
    %The solid angle of the surface surrounded by the entire trajectory will have a sudden change due to whether the evolution trajectory can cross the hemisphere as shown in Fig.6.
    %Whether the evolution trajectory can across the hemisphere will determine whether the phase transition occurs, 
    %This also gives explanation to the phenomenon in our previous simulation.
    %phase difference in our previous simulation and the phase transition when the PDL strength is around a certain value. 

    The PDL-induced non-reciprocal phases under different $\theta$ and $\eta$ are calculated and depicted in Fig.\ref{Fig:phase_diff}. In the case of weak PDL strength ($\eta > 0.96$), there is only a small fluctuation of the non-reciprocal phase with the change of $\theta$, as is depicted in Fig.\ref{Fig:Phase_Tran} in red. 
    %According to the previous analysis, this phenomenon is consistent with the situation where the weak measurement strength is low in the geometric phase theory. 
    In this case, the evolution trajectory of the polarization state (system state in the geometric phase theory) on the Bloch sphere is shown in Fig.\ref{Fig:Bloch_2}(a). We can find that the end point of the evolution trajectory cannot across the hemisphere. Therefore, after the final projection, the flux of the curved surface surrounded by the overall evolution trajectory yields a small value, corresponding to a small geometric phase. As the PDL strength increases, the end point of the evolution trajectory will finally crossed the hemisphere, and the solid angle of the trajectory will suddenly change, as shown in Fig.\ref{Fig:Bloch_3}(b). This happens at the point of $\eta = \eta_c \approx 0.96$ by setting $N=200$. in our numerical simulation, and the sudden changed phase curve is depicted in Fig.\ref{Fig:Phase_Tran} in blue. %In the end, the phase difference we found before can be understood as a geometric phase in classical light. 
    After all, this phase transition can be explained by the sudden change of the solid angle of closed trajectory on the Bloch sphere. 
    
    \section{Discussions}
    \subsection{Negative effects}
    
    In practice, the PDL in an optical fiber is usually weak, thus the non-reciprocal phase difference induced by PDL is also small. If the PDL-induced non-reciprocal phase is small comparing to the signal, this effect can be neglected.
    For instance, suppose the phase to be measured in the fiber Sagnac interferometer is in the order of $2\pi\times10^{-2}$. If we consider that
    the PDL-induced non-reciprocal phase being an order of magnitude lower than the signal is tolerable, then the overall PDL of the fiber loop $\Gamma_{dB}$ should be lower than 2.2dB. It is easily to be achieved in most of the optical fiber coils for normal applications. 
    
    However for a much smaller signal, the requirement for PDL strength will become more and more stringent. As is summarized in the following table, when the phase precision of the fiber Sagnac interferometer reaches to $2\pi\times10^{-6}$, the effect of PDL-induced non-reciprocity can not be neglected if $\Gamma_{dB}$ is larger than 0.0201dB. In this case, some parameters such as the radius of the fiber coil should be carefully considered when designing the fiber Sagnac interferometer\cite{Wang2007}.

    \begin{table}[]\centering
    \begin{tabular}{lllll}
    \hline
    \multicolumn{1}{c}{Phase precision} &$2\pi\times10^{-3}$ & $2\pi\times10^{-4}$ &$2\pi\times10^{-5}$ & $2\pi\times10^{-6}$ \\ \hline
    Maximum tolerable $\Gamma_{dB}$ &0.8002dB & 0.1603dB & 0.0501dB & 0.0201dB   \\ \hline
    \end{tabular}
    \caption{Maximum tolerable $\Gamma_{dB}$ under different sensitivity requirements.}
    \end{table}
    
    \subsection{Potential for sensing}
    The sudden phase change happens at the phase transition point provides a good opportunity for high sensitivity parameter estimation. Specifically, we can make the new-type non-reciprocal phase very close to the phase transition point by introducing a high strength PDL. At this point, a very slight change on the PDL will cause an abrupt change on the phase, which indicates an ultrahigh sensitivity.
    
    The extra PDL can be effectively introduced in many ways. For example, one can add an external pressure on a bent fiber from a different direction, as is shown in Fig.11. Here $\theta_{s}$ represents the angle between the directions of the pressure and the banding radius, but is not exactly equal to the one between the axes of extra birefringence and PDL, which requires more rigorous calculations\cite{Wang2007,bend1,bend2}. According to the previous analysis and results shown in Fig.5, we would need a PDL strength of about 42dB, which can in principle be realized by a 210 turns of single mode fiber coil with a radius of 9mm, according to the results presented in Ref.\cite{Wang2007}. This simple example is just for indicating the possibility. Many other possible ways, such as applying polarizing optical fiber or fiber Bragg grating, are worth to be explored for achieving the goal.

    \begin{figure}[h!]
    \centering
      \includegraphics[width=10cm]{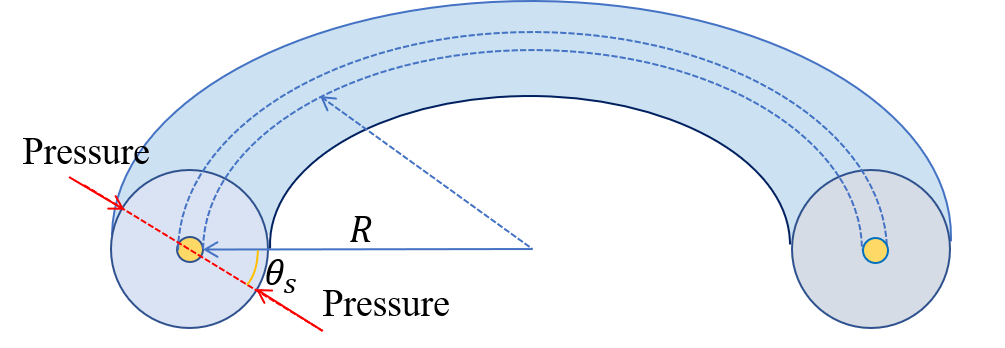} 
      \caption{Tilting the angle between the birefringence axis and the PDL axis by adding an external pressure on a bent fiber from a different direction. Here, R is the radius of curvature and $\theta_s$ is the angle between pressure direction and banding radius.}\label{Fig:application} 
    \end{figure}
    % At the same time, we are also aware that there are a large number of other non-reciprocal factors in the optical fiber ring. In a fiber optic Sagnac loop, there are many effects that will induce non-reciprocity, such as the power reflectance, the Kerr effect, the Faraday effect and so on. In particular, the reflected power from the interfaces of devices could be interfered with the signal and thus induce so-called amplitude-type noise. To avoid this problem, we can apply the method that widely used in fiber optic gyroscope, where a broadband light source takes the place of the narrow linewidth laser, in order to shorten the coherence length and significantly surpass the amplitude-type noise \cite{gyroscope}.
    Moreover, in practice there are many other physical effects that can induce non-reciprocal phase in a fiber Sagnac loop, such as the power reflectance, Kerr effect, Faraday effect and so on, which will presented as noises in our sensing application. For example, the reflected power from the interfaces of devices may introduce so-called amplitude-type noise, and a broadband light source is required for suppressing this noise\cite{gyroscope}. A more rigorous analysis on these kinds of noises and errors is an important issue in the future investigation.
    % For instance, it is possible to adjust the non-reciprocal geometric phase in the Sagnac loop near to the phase transition point by bending and squeezing the fiber. And then any small strain adding on the fiber will cause an small change of PDL strength, which will trigger a non-reciprocal phase transition that easily to be detected. This principle can be utilized to design new temperature or pressure sensors with ultra-high sensitivities. 
    
    % As is shown in Fig.\ref{Fig:Bloch_3}, the PDL-induced phase transition occurs when $\eta\approx 0.96$, corresponding to an overall PDL strength of $\Gamma_{dB} \approx 42dB$, therefore a special designed polarizing fiber or a fiber coil with very small radius\cite{Wang2007} is needed for designing such new-type sensors.
    %It is very rare in most fiber, unless it is a special polarizing fiber or is artificially coiled many times in a small radius. 
    %So another way of thinking, since the artificial bending and squeezing of the fiber can introduce birefringence and PDL, it is possible to adjust the geometric phase in the Sagnac loop to make it close to its own phase transition point. Any external stress that introduces additional PDL will cause the non-reciprocal phase to across the transition point. Then, this large phase change can be easily identified. We can use this principle to design a possible temperature or pressure sensitive trigger. 
    
    \section{Conclusion}
    
    In summary, we have found a new type of non-reciprocity in the optical fiber Sagnac interferometer,
    %This non-reciprocal phase 
    %irrelevant to the physical effects considered in previous studies,
    %has nothing to do with the path difference of the forward and reverse transmission light
    which is induced by the PDL (more precisely, the difference between the princial axes of PDL and birefringence). To explore the origin of this new non-reciprocity, we applied a recent proposed continuous-weak-measurement theory model and interpret the non-reciprocal phase as a PDL-induced geometric phase. Fortunately, this effect is so small that it can be neglected for most of the common applications using fiber Sagnac interferometers, unless the requirement for precision is extremely high.
    Moreover, the numerical simulation indicates that a phase transition phenomenon will happen when the PDL strength reaches a certain value, which implies a new possibility for designing new type of sensors with ultra-high sensitivity.
    
    \begin{backmatter}
    \bmsection{Funding}
    % This work was supported by the civil aerospace advance research project National Natural (D020403) and the Science Foundation of China (Grants No.62071298, No.61671287, No.61631014, and No.61901258).
    This work was supported by the civil aerospace advance research project (D020403) and the National Natural Science Foundation of China (Grants No.62071298, No.61671287, No.61631014, and 193 No.61901258).

    \bmsection{Disclosures}
    The authors declare that there are no conflicts of interest related to this article.
    
    \bmsection{Data availability}Data underlying the results presented in this paper are not publicly available at this time but may be obtained from the authors upon reasonable request.
    \end{backmatter}


\begin{thebibliography}{99}
    \bibitem{ref1} G. Sagnac. L'éther lumineux démontré par l'effet du vent relatif d'éther dans un nterféromètre en rotation uniforme. (C. R. Acad. Sci. Paris, 1913) t.157, pp. 708–710.
    \bibitem{ref2} D. H. Titterton, J. L. Weston. Strapdown Inertial Technology. (Amer Inst of Aeronautics, 1997).
    \bibitem{ref3} J. Blake, P. Tantaswadi and R. T. de Carvalho, In-line Sagnac interferometer current sensor, IEEE Transactions on Power Delivery, vol. 11, no. 1, pp. 116-121, Jan. 1996, doi: 10.1109/61.484007.
    \bibitem{ref4} L. Yundong, J. Xili, C. Hailiang, L. Jianshe, G. Ying, Z. Song, L. Hongyu, L. Shuguang. Highly sensitive temperature sensor based on sagnac interferometer using photonic crystal fiber with circular layout, Sensors and Actuators A: Physical, Volume 314, 2020, 112236, ISSN 0924-4247.
    \bibitem{ref5} D. Xinyong, H. Y. Tam, P. Shum. Temperature-insensitive strain sensor with polarization-maintaining photonic crystal fiber based Sagnac interferometer, Appl. Phys. Lett. 90, 151113 (2007).
    \bibitem{ref6} B. Culshaw. The optical fibre Sagnac interferometer: an overview of its principles and applications, 2005 Meas. Sci. Technol.
    \bibitem{ref7} G. Agrawal. Nonlinear Fiber Optics. (New York: Academic, 2001) pp. 467.
    \bibitem{ref8} Kurbatov, A.M., Kurbatov, R.A. Polarisation non-reciprocity cancelling in Sagnac fibre ring interferometer: an attempt of realistic study. Opt Quant Electron 51, 142 (2019).
    \bibitem{ref20} J. Hao, L. Zhou, Electromagnetic wave scatterings by anisotropic metamaterials: Generalized $4\times4$
    transfer-matrix method. Physical Review B, 2008, 77(9): 094201.
    \bibitem{ref19} T. Xu, F. Tang, W. Jing, H. Zhang, D. Jia, X. Zhang, G. Zhou, Y. Zhang, Distributed measurement of mode coupling in birefringent fibers with random polarization modes, Optica Applicata, Vol. XXXIX, No .1, 77- 90, 2009.
    \bibitem{ref18} G. Valentin, S. Kyrylo, W. Thomas, B. Andreas, R. Alessandro, G. Yuval, Topological transition in measurement-induced geometric phases. PNAS March 17, 2020 117 (11) 5706-5713.
    \bibitem{ref9} D. A. Jackson, R. Priest, A. Dandridge, and A. B. Tveten, Elimination of drift in a single-mode optical fiber interferometer using a piezoelectrically stretched coiled fiber, Appl. Opt. 19, 2926-2929 (1980)
    \bibitem{ref10} S. C. Rashleigh and R. Ulrich, High birefringence in tension-coiled single-mode fibers, Opt. Lett. 5, 354-356 (1980)
    \bibitem{ref11} A. M. Smith. Birefringence induced by bends and twists in single-mode optical fiber, Applied Optics, 1980, 19:2606-2611. D.
    \bibitem{ref12} C. Vessallo, Optical Waveguide Concepts. (Elsevier, 1991).
    \bibitem{ref13} Y. Zhou, W. Yu, H. Liu, J. Chen, P. Yang, L. She, C. Fang, J. Shao, Z. Guan, Z. Zhang, G. Feng, J. Yang, D. Chen. High-sensitive bending sensor based on a seven-core fiber. Optics Communications, Volume 483, 2021, 126617, ISSN 0030-4018.
    \bibitem{ref14} D. Marcuse. (1977), Loss Analysis of Single-Mode Fiber Splices. Bell System Technical Journal, 56: 703-718.
    \bibitem{ref15} D. Marcuse. Light Transmission Optics. New York, (Van Nostrand Reinhold Co, 1982).
    \bibitem{bend1} Makoto Tsubokawa, Tsunehito Higashi, and Yukiyasu Negishi, Mode couplings due to external forces distributed along a polarization-maintaining fiber: an evaluation, Appl. Opt. 27, 166-173 (1988).
    \bibitem{Wang2007} Q. Wang, G. Rajan, P. Wang, and G. Farrell, Polarization dependence of bend loss for a standard single mode fiber, Opt. Express 15, 4909-4920 (2007).
    \bibitem{bend2} P. Wang, G. Farrell, Y. Semenova, Q. Wang, A. M. Hatta, and G. Rajan Accurate theoretical prediction for single-mode fiber macrobending loss and bending induced polarization dependent loss, Proc. SPIE 7003, Optical Sensors 2008, 70031Y (28 April 2008);
    \bibitem{not_align} N. Brunner, A. Acín, D. Collins, N. Gisin, and Valerio, Scarani. Optical Telecom Networks as Weak Quantum Measurements with Postselection. Phys. Rev. Lett. 91, 180402(2003).
    \bibitem{ref16} S. Pancharatnam. Generalized theory of interference and its applications. Proc. Indian Acad. Sci. 44, 398–417 (1956).
    \bibitem{ref17} M. V. Berry. Quantal phase factors accompanying adiabatic changes. Proc. R. Soc, 1984, 392: 45-57.
    \bibitem{gyroscope} H. Lefevre, The Fiber-Optic Gyroscope (2nd Ed.), (Artech House , 2014).
    \end{thebibliography}
\end{document}